\documentclass[final,1p]{elsarticle}
\usepackage{color}
\usepackage{amssymb}
\usepackage{graphicx,color}
\usepackage[usenames,dvipsnames]{xcolor}
\usepackage[section]{placeins}
\usepackage{amsmath}
\usepackage{subfigure}
\usepackage[normalem]{ulem}
\usepackage{booktabs}
\usepackage{xcolor}
\usepackage{epsfig,graphics,color,graphicx,amsmath}

\colorlet{darkgreen}{green!50!black}
\colorlet{brightyellow}{yellow!75!red}
\colorlet{orange}{red!50!yellow}
\colorlet{darkblue}{blue!60!black}
\colorlet{darkred}{red!80!black}

\def\psla{  \slash \!  \!\!}
\def\Psla{   \slash \! \! \!\!}
\def\be{\begin{eqnarray} &&}
\def\nonu{\nonumber \\ &&}

\def\ee{\end{eqnarray}}
\newcommand{\bfm}[1] {\mbox{\boldmath{$#1$}}} 

\usepackage{soul}

\usepackage{mathtools}
\usepackage{mathrsfs}
\usepackage{bm}
\usepackage{relsize}
\usepackage{indentfirst}

\usepackage{xcolor}

\journal{Physics Letters B}
\begin{document}
\begin{frontmatter}

\title{Pion electromagnetic form factor with Minkowskian dynamics}
\author[1]{E.~Ydrefors}
\author[1]{W.~de Paula}
\author[1,3,4]{J.H.~Alvarenga Nogueira}
\author[1]{T.~Frederico\corref{cor1}}
\ead{tobias@ita.br}
\author[4]{G.~Salm\`{e}}
\address[1]{Instituto Tecnol\'ogico de Aeron\'autica,  DCTA, 12228-900 S\~ao Jos\'e dos Campos,~Brazil}
\address[3]{Phys. Dept. Rome Univ. ''La Sapienza'', Piazzale A.~Moro 2 
- 00185 Roma, Italy}
\address[4]{INFN, Sezione di Roma, Piazzale A.~Moro 2 - 00185 Roma, Italy}
\cortext[cor1]{Corresponding author}

\date{\today}

\begin{abstract}
 The pion electromagnetic form factor has been calculated for the first time from the solutions of the Bethe-Salpeter
equation, obtained directly in Minkowski space by using the Nakanishi integral representation of the Bethe-Salpeter
amplitude. The  one-gluon exchange kernel contains as inputs the quark and gluon masses, as well as a 
scale parameter featuring the extended quark-gluon vertex.  The range of variability of these parameters is suggested  by lattice 
calculations. After presenting
  a very  successful comparison with   the existing data in the whole range of the momentum transfer, we show  how inserting the light-front (LF) formalism
in our approach allows to achieve further interesting results, like the evaluation of the LF valence form factor and 
 a first investigation of the end-points effects.
\end{abstract}
\end{frontmatter}

%\section{Introduction}\label{Sec:intr}

The pion  is intimately associated with our understanding of the mass generation of the visible matter in the 
universe~\cite{horn2016pion}, and therefore the detailed study of  its structure has necessarily
a central role in hadron physics.
The  non perturbative dynamics of 
Quantum Chromodynamics (QCD) is enclosed in the hadron structure, and its description in Minkowski space represents one of the challenges faced by the theory. 
The current experimental and theoretical efforts  to achieve a 3D imaging  of hadrons, which is one of 
the goals of the future Electron  Ion Collider (see e.g. Ref.~\cite{Accardi:2012qut}), 
demand information from the measurements of 
 various observables, as well as from different theoretical tools.
  Although a
huge amount of progress  has been made in Euclidean treatments of QCD either by discrete 
(see, e.g., Ref.~\cite{Oehm19}) or continuum
methods (see, e.g., Refs \cite{CloPPNP14,Eichmann:2016yit}),   Minkowskian phenomenological approaches 
to solve QCD  could   non trivially contribute to   the  cooperative efforts at investigating the dynamics
inside hadrons. Examples of practical
methods to extract Minkowski space observables from Euclidean treatments of QCD are based on:  (i) computing  moments of 
 hadron parton momentum 
distributions~ \cite{Oehm19,Shi:2018zqd,Bednar:2018mtf},  (ii) resorting to a  boost 
of the hadron to the infinite 
momentum frame on the Lattice~\cite{XJIPRL13,Sufian:2020vzb}, which has its inherent difficulties~
\cite{Rossi:2017muf},  (iii) adopting both  Ioffe-time and transverse coordinates \cite{Radyushkin:2017cyf},
 (iv)  using integral 
representations to bridge Euclidean and Minkowskian amplitudes~\cite{LeiPRL13,carbonell2017euclidean}  or (v)  exploiting  
 the 
Basis light-front  quantization~\cite{Lan:2019vui}.

The Minkowskian approach  we are pursuing 
to describe the pion \cite{dePaula:2021} brings the relevant dressing 
scales from the non perturbative QCD, namely the constituent quark and gluon masses, and an extended quark-gluon vertex. However, we do not yet solve self-consistently the
Bethe-Salpeter equation (BSE) and the truncated Dyson-Schwinger  
equations (DSEs), which would allow  the dynamical chiral symmetry breaking,  as done by the four-dimensional continuum approaches in Euclidean space  
(see e.g. Ref. \cite{Eichmann:2016yit}).
In particular,
 they use a suitable kernel  to cope with both ultraviolet  and
 infrared (IR) regions,
  needed for realistically  implementing   both   chiral symmetry breaking and confinement. It is understood that such features are very important for developing  QCD-inspired dynamical models,  and should be conveyed  in  Minkowskian  approaches 
that aim to be applied to the pion phenomenology. Among the pion observables,
 the electromagnetic (em) form factor (FF) plays a
relevant role for accessing the inner pion structure, since it is related to the charge density in the so-called impact parameter
space (see, e.g., Refs.~\cite{Soper:1976jc,Brodsky:2007hb,Miller:2007uy}),  as well
as to the unpolarized generalized parton distribution (see, e.g., Ref.~\cite{Frederico:2009fk}). Therefore, efforts to 
improve the
dynamical descriptions of the pion in the physical space in synergy with   Euclidean   phenomenological approaches (see,
e.g., Ref.~\cite{ChanPRL13} for a  description of the em FF in continuum QCD)
become highly desirable.

 Within our approach \cite{dePaula:2020qna}, based on the 4D BSE in Minkowski space and the 
 Nakanishi integral representation (NIR) of the Bethe-Salpeter amplitude,  we have evaluated the em FF of the pion. 
  The interaction kernel is taken in the ladder approximation,  with three input  parameters:  the constituent-quark 
  and gluon masses, and  a scale parameter
featuring the extended quark-gluon vertex. 
The actual
values of the three quantities are inferred 
 from lattice QCD (LQCD) calculations.
  In addition, we note that the  ladder kernel  should be considered   a reliable approximation to 
  compute the pion bound state,
   as suggested by the suppression of the non-planar contributions for $N_c=3$  within the BS approach in a scalar QCD model~\cite{Nogueira:2017pmj}. 
    It should be pointed out that
the pion BS amplitude, in spite of  its simple dependence upon two fermionic fields, has an infinite 
 content of Fock states (the use of the Fock space allows to recover a probabilistic language) and the formal link  
with the valence wave function, the amplitude of the Fock component of the pion state with the lowest number of constituents, 
is provided by the so-called LF projection. This formal procedure amounts to  the integration of the BS amplitude over the minus component of the relative 4-momentum (see, e.g., Refs. \cite{FSV1,Sales:2001gk,Marinho:2008pe}).
 Within the proposed framework, the pion valence wave function  contributes only with 
 70\%~\cite{dePaula:2020qna}
 of the normalization, and consequently to the charge  (in impulse approximation).  Therefore, 
  this component  is expected to miss a significant amount of the FF strength 
 in the low and medium range 
 (to be determined) of the square momentum transfer, $Q^2$, while  
 at large
 momentum transfers only its contribution should survive \cite{LepPLB1979,LepPRD1980}. Noteworthy, 
  the  above mentioned deficiency highlights  in turn the  role of the higher Fock components in the same range of $Q^2$.
  
  Summarizing,    our approach, carried out  in Minkowski space and 
  genuinely dynamical, has the advantage (i) to allow a decomposition in terms of valence 
and non valence LF Fock components,
along with the further decomposition of the valence contribution in terms 
of the pion spin configurations, and (ii) to pave the way for investigating   the transition to the regime  of valence dominance, so that the  impact of the 
  higher Fock states 
  on the low $Q^2$ region can be assessed.  This analysis should be relevant, in view of the ongoing 
  experimental and theoretical efforts for understanding the valence structure of the pion 
  (see, e.g., the recent Electron Ion Collider Yellow Report \cite{khalek2021science}).

%%%%%%%%%%%%%%%%%%

{\it The pion BSE and  the Nakanishi integral representation.} We shortly recall our Minkowski space formalism for describing the pion as a $q\bar q$ bound state. One can find a more 
detailed treatment  in Refs.~\cite{dePaula:2016oct,dePaula:2017ikc} and  a wide numerical analysis of
both static and dynamical quantities  in Ref.~\cite{dePaula:2020qna}.

The fermion-antifermion BSE in the ladder approximation reads 
\begin{equation}\label{Eq:BSE}
  \Phi(k; P) = S\bigl(k+\tfrac{P}{2}\bigr) \int \frac{d^4k'}{(2\pi)^4}S^{\mu\nu}(q)\Gamma_{\mu}(q)
  \Phi(k';P)\widehat\Gamma_{\nu}(q)S\bigl(k-\tfrac{P}{2}\bigr),
\end{equation}
where $p_{1(2)} = \tfrac{P}{2} \pm k$, with 
$p^2_{1(2)}\neq m^2$, are the momenta of the two off-mass-shell particles.
 $P = p_1 + p_2$ is the total momentum, with $P^2 = M^2$ the squared bound-state mass,  $k=\tfrac{1}{2}(p_1 - p_2)$ is the relative 
four-momentum  and   $q = k - k'$  is the momentum transfer. Moreover, one has 
$\widehat\Gamma_{\nu}(q)=C~\Gamma_{\nu}(q)
~C^{-1}$,
where  $C=i\gamma^2\gamma^0$  is the  charge-conjugation operator. 
In Eq.~\eqref{Eq:BSE},   the fermion propagator,  the gluon 
propagator in the Feynman gauge  and  
the   quark-gluon  vertex,  dressed through a simple form factor, 
are 
\be
S(P) =  ~{i \over \psla{P} - m + i\epsilon}~, \qquad 
    S^{\mu\nu}(q) = -i \frac{g^{\mu\nu}}{q^2 - \mu^2 + i\epsilon}~,\qquad~
    \Gamma^\mu= i g \frac{\mu^2 - \Lambda^2}{q^2 - \Lambda^2 + i\epsilon}\gamma^\mu~,
\ee
where $g$ is the coupling constant, $\mu$ the mass of the exchanged boson 
and $\Lambda$ is a scale parameter
chosen to suitably model the  color distribution  in the
 interaction vertex of the 
dressed constituents. 

The BS amplitude, $\Phi(k; P)$, can be decomposed as
\begin{equation}\label{BS_decomp}
  \Phi(k;P) =  S_1(k;P) \phi_1(k;P)+S_2(k;P) \phi_2(k;P)+S_3(k;P) \phi_3(k;P)+S_4(k;P) \phi_4(k;P)~~,
\end{equation}
 where the $\phi_i$'s are scalar functions, and $S_i$'s are 
 Dirac structures  given by
\begin{equation}
%  \begin{aligned}
    S_1(k;P) = \gamma_5, \,\, S_2(k;P) = \frac{\Psla{P}}{M}{\gamma_5}, \,\,
    S_3(k;P) = \frac{k\cdot P}{M^3}\Psla{P}\gamma_5 - 
    \frac{1}{M}\psla{k}\gamma_5, \,\, S_4(k;P) =
    \frac{i}{M^2}\sigma^{\mu\nu}P_\mu k_\nu \gamma_5\, .
%    \end{aligned}
\end{equation}
 Notice that  the anti-commutation rules of the fermionic fields
 impose   the functions  $\phi_i$  must be 
 even for $i=1,2,4$ and odd for $i=3$, under the change 
 $k \rightarrow - k$.

The  scalar functions $\phi_i(k;P)$ in \eqref{BS_decomp} can be written in terms of the NIR:
\begin{equation}
  \label{Eq:NIR}
  \phi_i(k; P) = \int_{-1}^1 dz' \int_0^\infty d\gamma'\frac{g_i(\gamma',z';\kappa^2)}{[k^2 + z'(P\cdot k) - \gamma' - \kappa^2 + i \epsilon]^3}, 
\end{equation}
where  $\kappa^2 = m^2 -{M^2/4}$, and  $g_i(\gamma',z';\kappa^2)$ are the 
 Nakanishi weight 
functions (NWFs), that   are real and
 assumed   to be unique, following    the uniqueness theorem from Ref.~\cite{Nakanishi:1971}. 
 Noteworthy, the NWFs enclose all 
the non perturbative dynamical information included in the BS interaction kernel. 

By inserting Eqs.~\eqref{BS_decomp} and \eqref{Eq:NIR} in the BSE,
Eq. \eqref{Eq:BSE}, and subsequently performing a LF projection, 
one can formally transform the BSE 
into a  
 coupled system of integral equations for the NWFs (see details in
Ref.~\cite{dePaula:2017ikc}).

\begin{figure}[htp]
  \centering
  \includegraphics[width=15.0cm]{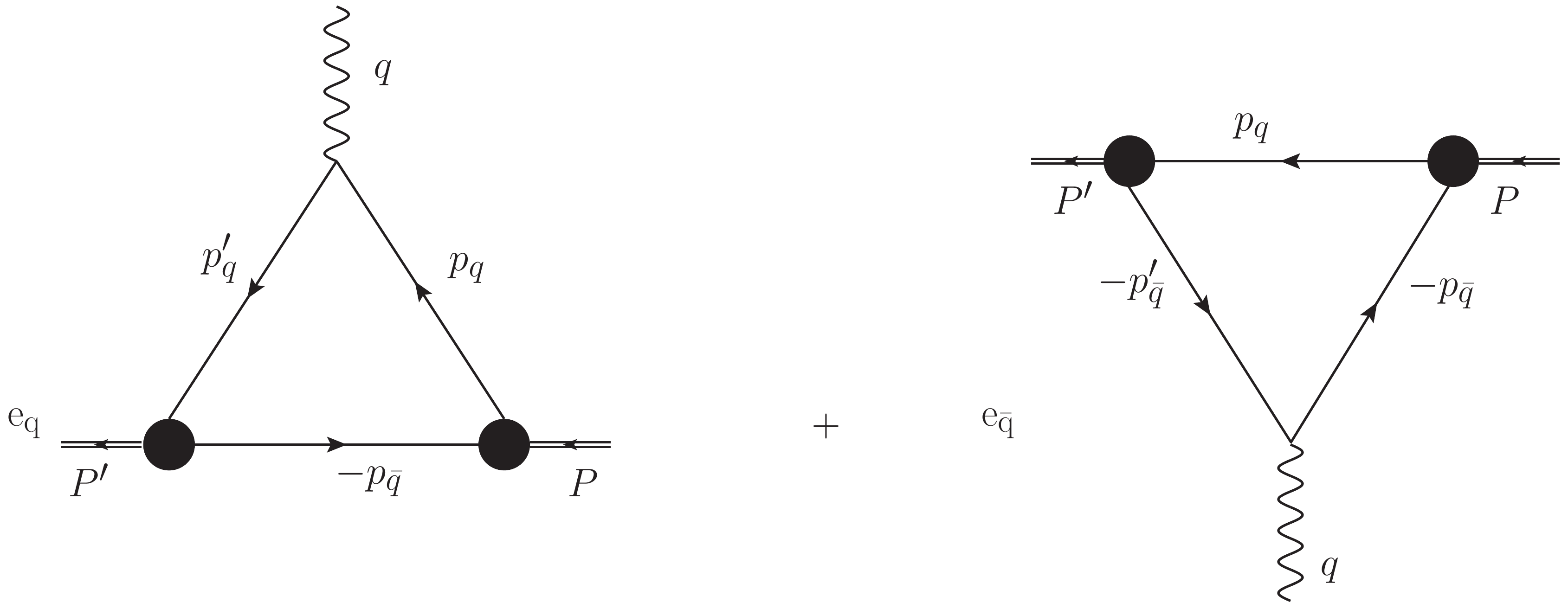}
  \caption{Diagrammatic  representation of  the pion electromagnetic form factor 
  in one-loop approximation (cf Eq.~\eqref{Eq:FF2}). The  full dots
  represent the vertex  functions for the initial and final bound states, respectively. Each vertex function, Eq. \eqref{Vqpion}, contains the non perturbative dynamics and 
  can be expressed in terms of the BS amplitude. The charge of the quark
  (anti-quark) is indicated by $e_{q(\bar q)}$.}
\label{fig:FFdiag}
\end{figure}

{ \it The covariant expression of the electromagnetic form factor.} To evaluate the elastic em FF of the pion, it has been adopted the quantum-field
 formalism
  proposed by Mandelstam in  Ref. \cite{Mandelstam:1955sd}, and
 originally  applied to the deuteron. 
 The leading 
 idea was to take benefit of  the prevalence of a bound-state pole  in a suitable 
 vacuum correlation function, where both the  em current operator and the fermionic
 fields needed for extracting   bound-state contributions (for  initial and final
 two-body states) are present.
 Eventually,  one is able to bring out  
 the em 
 current  matrix elements between the relevant 
 bound-states, obtaining an expression in terms of the pion BS amplitudes.  
 Remarkably, the pion BS amplitude
  contains contributions not only from a two-fermion state, but also from 
  an infinite set of
  states with two fermions and any number of gluons, that in the present 
  study correspond 
    to
   the infinite set of ladder
  exchanges necessary to reconstruct the bound-state pole.
  
   By adopting the  impulse approximation, i.e. 
 a bare quark-photon vertex $i ~\gamma^\mu$,    the elastic  em FF  of a charged pion is 
pictorially represented   {\it \'a la Feynman } in Fig.~\ref{fig:FFdiag},  by the sum of two 
 triangle (one-loop) diagrams, weighted by the  charge of the quark interacting
  with the
 virtual photon. Differently, the vertex between  constituents and  
bound-system  is a dressed one, described 
 in terms of the BS amplitude by (see Fig.~\ref{fig:FFdiag})
 \be
 {\cal V}(k,P)= S^{-1}(p_q)~\Phi (k;P)~S^{-1}(p_{\bar q}),
 \label{Vqpion}\ee
 where $p_q$ and $p_{\bar q}$ are the quark and anti-quark momenta, respectively, and $S^{-1}(p_{\bar q})=-i~(-\psla{p}_{\bar q} - m)$.
  In the case  of a positive pion, within a $SU(2)$ framework, one has 
  two contributions, 
    where the virtual photon is absorbed by a $u$ quark or by a $\bar d$  one. 
  Due to the underlying symmetries, both terms are equal (when taking the absolute value of
the quark charge) 
 so that  
     the elastic FF is given by 
\begin{equation}
  \label{Eq:FF2}
  F(Q^2) = -i~{ N_c\over  M^2~(1+\tau)}
  \int \frac{d^4 p_{\bar q}}{(2\pi)^4}~\text{Tr}
  \left[(-\psla{p}_{\bar q} - m)\bar
  {\Phi} (k';P^\prime) (\Psla P +\Psla P')~ \Phi(k;P)
  \right],
\end{equation}
where $N_c$ is the number of colors, $\bar
  {\Phi} (k';P^\prime)$ is the BS-conjugated amplitude (see the explicit
  expression in Ref. \cite{dePaula:2020qna}), 
$Q^2 =-q^2= -(P^\prime-P)^2=-(p^\prime_q-p_q)^2$, $(P + P^\prime)^2= 4M^2 (1+\tau)$ with $\tau=Q^2/4M^2$ 
  and the initial and final relative momenta are 
  \be  
   k = \frac{P}{2} - p_{\bar q}=p_q-\frac{P}{2}~~, \quad \quad 
k' = \frac{P^\prime}{2} - p_{\bar q}=p^\prime_q-\frac{P^\prime}{2}=k+{q\over 2}.
\ee
For $q\to 0$,  one  has to obtain  the charge normalization, $F(0)=1$, 
that in ladder 
approximation   
is equivalent  to the BS normalization (see Ref. \cite{dePaula:2020qna}).
 
 By  inserting
  the decomposition \eqref{BS_decomp} and the NIR of the scalar functions
  $\phi_i(k;P)$ in Eq. \eqref{Eq:FF2},  one gets
\be
\label{Eq:FF2b}
F(Q^2) =~-i {N_c\over 4M^2 ~(1+\tau)}~\sum_{i,j=1}^4 \int \frac{d^4 k}{(2\pi)^4}\int_0^\infty d\gamma \int_{-1}^1dz
     \int_0^\infty d\gamma'\int_{-1}^1 dz'~g_j(\gamma, z)g_i(\gamma',z')\nonu \times ~
     {\text{Tr}\Bigl[(\psla k -\Psla P/2-m)~
{ \gamma^0
~S_j^\dagger(k';P^\prime)~\gamma^0}
  (\Psla{P}+\Psla{P}^\prime)~S_i(k;P)\Bigr]
     \over [\gamma + \kappa^2 - k^2 -
    (P\cdot k)z - i\epsilon]^3~~[\gamma' + \kappa^2  - k'^2 -
    (P^\prime\cdot k')z' - i\epsilon]^3}~~,
\ee
After  performing
both
 traces and  4D integration,  
one gets the final expression
   for the elastic em FF, viz.
\be
  \label{Eq:FF_final}
  F(Q^2) =  \frac{N_c}{32\pi^2}\sum_{i,j=1}\int_0^{\infty}d\gamma \int_{-1}^1 dz 
 \int_0^{\infty}d\gamma'\int_{-1}^1 dz' ~g_j(\gamma,z)~g_i(\gamma', z')
  \nonu \times ~\int_0^1 dy~ y^2(1-y)^2 ~\frac{\tilde c_{ij}(y,\gamma,z;\gamma',z';Q^2)}
  {\Bigl[\kappa^2 + \gamma' \, (1-y) + \gamma \, y +  \alpha^2 \, M^2 \, 
+ {Q^2\over2} \, y \, (1+z) \, \beta\Bigr]^4}~~,
\ee
where the coefficients $\tilde c_{ij}$, $\alpha$ and $\beta$ are given in 
\ref{App:4Di}.

%%%%%%%%%%%%%%%%%%%%%%%%%%%%%%%%%
 Besides  the covariant expression, one could equivalently analyze 
the pion em FF within the    LF framework,  and obtain the
so-called 
Drell-Yan expression of the elastic FF (see, e.g., Ref.~\cite{Brodsky:1997de}).
  As is well-known, within the LF framework, 
  once  a tiny mass is assigned to the  exchanged boson, one can introduce a meaningful Fock-expansion of a bound-system state~\cite{Brodsky:1997de}.  
  In particular, the valence wave function can be extracted from  the BS amplitude through 
  the LF projection 
  onto the null-plane, $x^+=0$ (see Refs.~\cite{dePaula:2020qna,FSV1,Sales:2001gk,Marinho:2008pe} for more details).
  
In the Drell-Yan frame ($q^+=0$), the em current operator component, $J^+=J^0+J^3$, 
 is diagonal in the LF Fock-space~\cite{Brodsky:1997de}, and 
 one  can profitably decompose  
 the pion FF 
 into  two main contributions: (i) the LF valence contribution and (ii) the non valence one, 
 generated  by the higher Fock-components 
of the pion state, viz (see also Ref. \cite{Frederico:2009fk})
\begin{equation}\label{decomp}
F(Q^2)=\sum_{n=2}^\infty F_n(Q^2) =F_{\rm val}(Q^2)+F_{\rm nval}(Q^2)\, ,
\end{equation}
where $F_n(Q^2)$ represents the contribution of the $n-$th Fock component of the
pion state, $F_{\rm val}(Q^2)=F_{n=2}(Q^2)$ is the 
LF valence contribution  normalized as $F_{\rm val}(0)=P_{\rm val}$, with $P_{\rm val}$ the valence probability (cf. Ref.
\cite{dePaula:2020qna}), and $F_{nval}(Q^2)$ the non valence contribution with normalization $1 -P_{\rm val}$.
Hence, once we are able to compute  both the full FF and  the
  LF valence 
 contribution, 
the non valence part  can be obtained by using  Eq.~\eqref{decomp}, so that the role of 
higher Fock-components in generating various features of the FF can be assessed, 
e.g., by evaluating
the non valence contribution to the charge radius.

%%%%%%%%%%%%%%%%%%%%
 
{\it The  LF valence contribution to the em  form factor.} The valence contribution to the FF is obtained from the   matrix
elements of the component $\gamma^+$ of the  bare em current, between the initial and final LF valence states.  
In a Drell-Yan frame, where
 ${\bf P}_\perp + {\bf P}_\perp^\prime=0$,  
one  obtains
the following expression 
 (see, e.g., Refs.~\cite{Frederico:2009fk,Sales:2001gk,Marinho:2008pe,MezragFBS})
 \begin{equation}
  \label{Eq:val_FF}
    F_{\text{\rm val}}(Q^2) = F^{S=0}_{\text{\rm val}}(Q^2)+F^{S=1}_{\text{\rm val}}(Q^2)
  %\nonu
  =  \frac{N_c}{16\pi^3}\int d{\bfm  \kappa}_\perp 
    \int_{-1}^1 dz 
   \Bigl[\psi^*_{\uparrow\downarrow}(\gamma',z)\psi_{\uparrow\downarrow}(\gamma,z)
+ \frac{{\bfm \kappa}_\perp \cdot {\bfm \kappa}^\prime_\perp}{\sqrt{\gamma\gamma'}}
 \psi^*_{\uparrow\uparrow}(\gamma',z)\psi_{\uparrow\uparrow}(\gamma,z) \Bigr] \, , 
\end{equation}
 where  $Q^2 = | {\bf q}_\perp|^2$,    
$\gamma = |{\bfm \kappa}_\perp|^2$,    
 $\gamma'= |{\bfm \kappa}'_\perp|^2$
 %=\gamma +{(1+z)^2\over 4} Q^2 + (1+z)~\sqrt{\gamma~Q^2}~\cos\theta$ 
 and
 ${\bfm \kappa}'_\perp={\bfm \kappa}_\perp+ {1+z\over 2}~ {\bf q}_{\perp}$.
 Notice that  $z=(1-\xi)/2$ with  the longitudinal-momentum fraction $\xi=p^+_q/P^+$, and hence for $z=-1$, i.e. $\xi =1$, 
 the initial and final  intrinsic momenta  are collinear. 
 
 The two spin 
components 
$\psi_{\uparrow\downarrow}$ and $ \psi_{\uparrow\uparrow}$, 
corresponding to the anti-aligned ($S=0$) and aligned ($S=1$) spin
configurations, are proper integrals of the NWFs (see details in Ref. 
\cite{dePaula:2020qna}).
 
   In order to study the asymptotic behavior of the LF valence FF, 
  we first observe that  the rightmost contribution to Eq.~\eqref{Eq:val_FF} can be disregarded (checked also numerically) 
  since 
 (i) ${\bfm \kappa}_\perp \cdot {\bfm \kappa}^\prime_\perp$  vanishes for $\gamma\to 0$, precisely where the functions 
 $\psi_i$ are
  bigger, and (ii)  the overall weight of the $S=1$ configuration  is smaller with respect to the  $S=0$ one
  (cf. the calculated probabilities in Ref. \cite{dePaula:2020qna}).
 Moreover, the  fall-off of the $\psi_{\uparrow\downarrow}$  sets an effective range of $\gamma$
 where the amplitudes yield sizable contributions. In conclusion, by disregarding the $S=1$ contribution and
 approximating $ \psi_{\uparrow\downarrow}(\gamma',z)\sim 
 \psi_{\uparrow\downarrow}\left((1+z)^2Q^2/4,z\right)$
 for large $Q^2$ one gets for the asymptotic em FF, namely $F^{(a)}_{val}(Q^2)$, 
 \be
  \label{Eq:val_FF2}
    F_{\text{\rm val}}(Q^2)|_{Q^2\to\infty}\sim~ {F^{(a)}_{\text{\rm val}}(Q^2)=}
    \frac{N_c}{16\pi^2}  \int_{-1}^{1}dz \,\psi_{\uparrow\downarrow}\left(\frac{(1+z)^2}{4}Q^2,z\right)
\,   \int_0^\infty d\gamma\, \psi_{\uparrow\downarrow}(\gamma,z)\, ,
\ee
where the last integral over $\psi_{\uparrow\downarrow}(\gamma,z)$ provides 
the corresponding {\em distribution 
amplitude}.

The asymptotic LF valence FF represents a physically motivated reference line for the valence  FF, like  
 the QCD asymptotic expression  given in Ref.~\cite{LepPLB1979} does  for the full FF.
 In particular, the asymptotic QCD FF, based on an asymptotic
approximation of the
anti-aligned spin component of the pion wave function, reads
\be
 Q^2F_{\text{asy}}(Q^2)=8\pi \alpha_s(Q^2) f^2_\pi \, ,
\label{pQCD}\ee
where $\alpha_s(Q^2)$ comes from \cite{Zyla:2020zbs}  (notice that  Eq. (4.4) of Ref. 
\cite{LepPLB1979} has a different definition of $f_\pi$) and the decay constant
is related to the anti-aligned component (see, e.g., Ref.\cite{dePaula:2020qna}).
A nice feature of the approximate  formula in Eq. \eqref{Eq:val_FF2}  is given by the
 possibility to understand
 that the integral in $z$ receives contributions, for large $Q^2$,  from the region close to $z=-1$ ($\xi=1$), when we
 consider a quark ($z=1$, i.e.~$\xi=0$ for an anti-quark), 
 since
  the amplitude  $\psi_{\uparrow\downarrow}(\gamma',z)$ is sizable for
small values of $\gamma'=(1+z)^2~Q^2/4$. On the other hand, the amplitude is  damped at the end-points, $z=\pm 1$. 
Therefore, from the competition of the two effects, one should expect that the integrand in $z$ of the total valence FF,  Eq.~\eqref{Eq:val_FF},  should be peaked close to 
$z= -1$ 
for large momentum transfers.
This will be illustrated
when presenting our numerical results. 

%\centerline{ Table}
\begin{table}[thb] 
\begin{center}
\begin{tabular}{c c c c c c c c c c}
\hline
Set  & $m$  & $B/m$ & $\mu/m$ & $\Lambda/m$ &  $P_{\rm val}$ &$f_\pi $ & $r_\pi$ (fm)  & $r_{\rm val}$ (fm) & $r_{nval}$ (fm)  \\
\hline
I  & 255  & 1.45  & 2.5  & 1.2  &  0.70 & 130 & 0.663 & 0.710 & 0.538 \\
II  & 215  & 1.35  & 2    & 1    &  0.67 & 98  & 0.835 & 0.895 & 0.703\\
\hline
\end{tabular}
\caption{ \label{tableradius}   The pion model, with $M=m_\pi=140$ MeV, for two parameter sets and the corresponding
(i)  valence probabilities
 $P_{\rm val}$, (ii) decay constants  $f_\pi$ in MeV, (iii) pion charge radii,  and (iv) their decomposition in  
valence and non-valence contributions (see Eq. \eqref{raddeco}).  
The experimental  values for the pion charge radius and $f_\pi$ are 
$r^{PDG}_\pi=0.659\pm0.004$ fm,
 and 
 $f^{PDG}_\pi=130.50 (1)(3)(13)$ MeV, respectively (both from   Ref. \cite{Zyla:2020zbs}). }
\end{center}

\end{table}

%\section {Numerical results} \label{Sec:Res}
{\it Numerical results.} The solutions of  the coupled system for the NWFs were obtained after discretization that follows 
by expanding  $g_i(\gamma, z)$ onto a bi-orthogonal basis, i.e.~Laguerre polynomials 
for the non-compact variable, $\gamma$, and Gegenbauer polynomials for the
compact one, $z$. In this way,   one reduces the BSE into
 a generalized eigenvalue problem, where 
the squared coupling constant $g^2$ is the eigenvalue  (see, e.g., Ref.~\cite{dePaula:2017ikc}). 
As inputs, we have  used the binding energy 
$B=2m -M$, the exchanged-boson mass, 
$\mu$, and  the scale parameter, 
$\Lambda$, present in the extended quark-gluon vertex. 

The  adopted   parameter sets, shown 
in Table \ref{tableradius}, provide  values of $f_\pi$ equal to   $130$ and $98$   MeV,
respectively (for a wider discussion and details, see Ref. \cite{dePaula:2020qna}).
 Notice that: (i)  the gluon mass $\mu$ is chosen as  $637$ and $430$  MeV,  with the 
first  value   
suggested by   LQCD results 
for the dressing function in the IR region, ranging in the interval 
 723-648 MeV in the Landau gauge~\cite{OLJPG11}; (ii) the constituent 
quark mass $m$ is around 250 MeV, in order
 to be close 
to the IR values of the running mass in LQCD, 
varying in the interval 280-240 MeV~\cite{ParPRD06} for momenta below $\sim 2\Lambda_{QCD}$; (iii) the parameter $\Lambda$  is 
taken of the order of $\Lambda_{QCD}$~\cite{Oliveira:2018ukh, Oliveira:2020yac}.
 The set II is chosen
for illustrating the crucial  role of the correct reproduction 
of $f_\pi$, even with  model parameters quite close to set I.

 In Table~\ref{tableradius}, it is also shown the  pion  charge radius and its decomposition in valence and non valence contributions. It is observed that, in correspondence to the experimental
value of the decay constant, $f_\pi=130$ MeV, we  get  $r_\pi=0.663$~fm, in fair agreement with the 
most recent PDG value, i.e.~$r^{PDG}_\pi=0.659\pm0.004$ fm~ \cite{Zyla:2020zbs}. 
This is expected from the correlation between the two quantities (see Ref.~\cite{tarrach1979meson}).
The  decomposition of the charge radius, that can yield  valuable insights on the role of the higher Fock states, is straightforwardly  obtained by using (i) the definition 
$r^2_\pi=-6~d F_\pi(Q^2)/{dQ^2}\Bigr|_{Q^2=0}$, (ii) the decomposition given in Eq.~(\ref{decomp})
 and (iii) the normalization of the valence and non valence em FFs. Explicitly, it reads  
 \be
r^2_\pi=P_{\rm val}\, r^2_{\rm val}+(1-P_{\rm val})\,r^2_{nval}\quad \text{with}\quad P_{\rm val(nval)}\,r^2_{\rm val(nval)}=-6~d F_{\rm val(nval)}(Q^2)/{dQ^2}\Bigr|_{Q^2=0}.
\label{raddeco}
\ee

   Table \ref{tableradius} illustrates that the higher Fock components generate a
charge distribution more compact than the one   from the LF valence wave function, 
i.e.~$r_{\rm val}> r_{nval}$. 
The physical interpretation of this feature is quite natural, once we recall that the 
higher Fock components of the pion
have  larger
 virtualities,  leading to  very short lifetimes.  Hence, there is no room for the quarks     
to fly  too far from  the pion center  of mass.

\begin{figure}[thp]
  \centering
  \includegraphics[scale=0.67, angle=0]{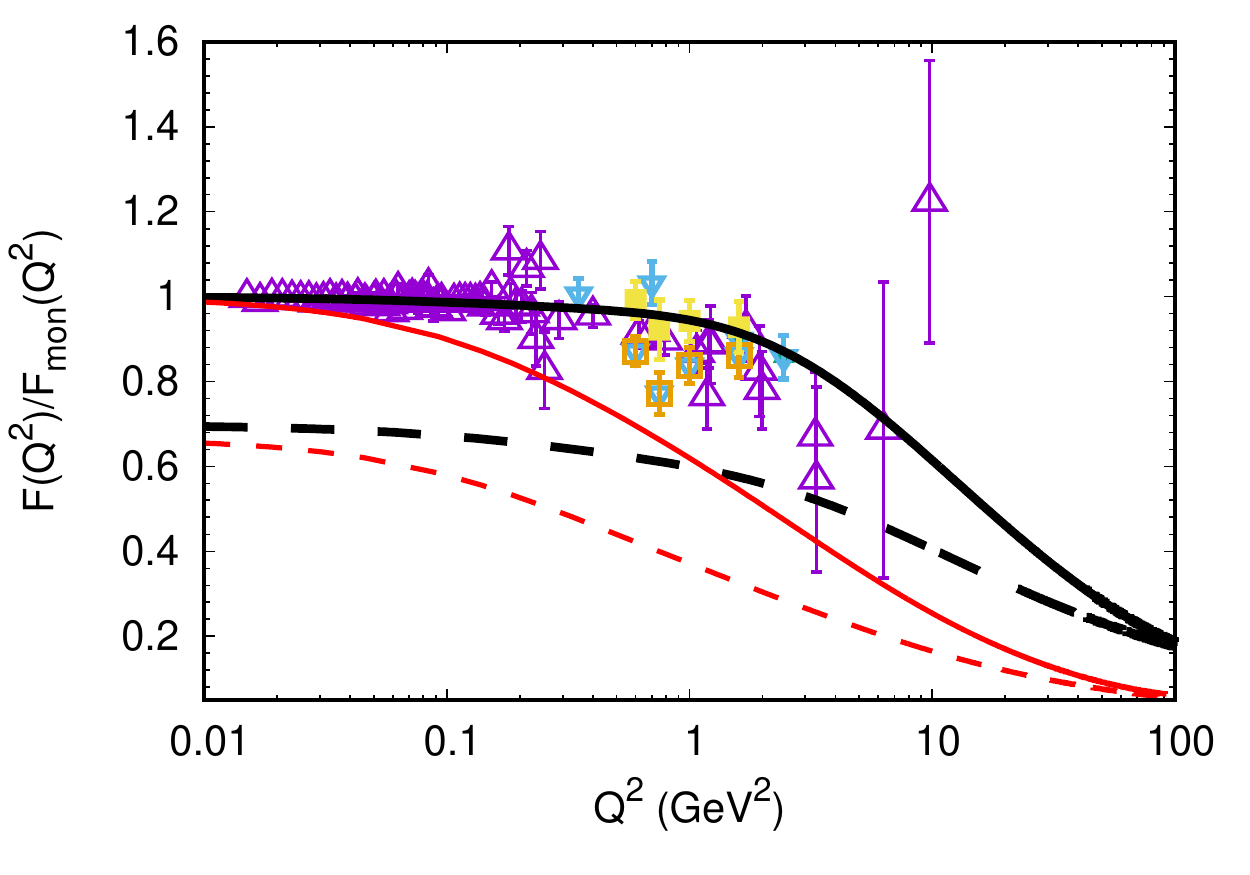} \hspace{-0.4 cm}
   \includegraphics[scale=0.67, angle=0]{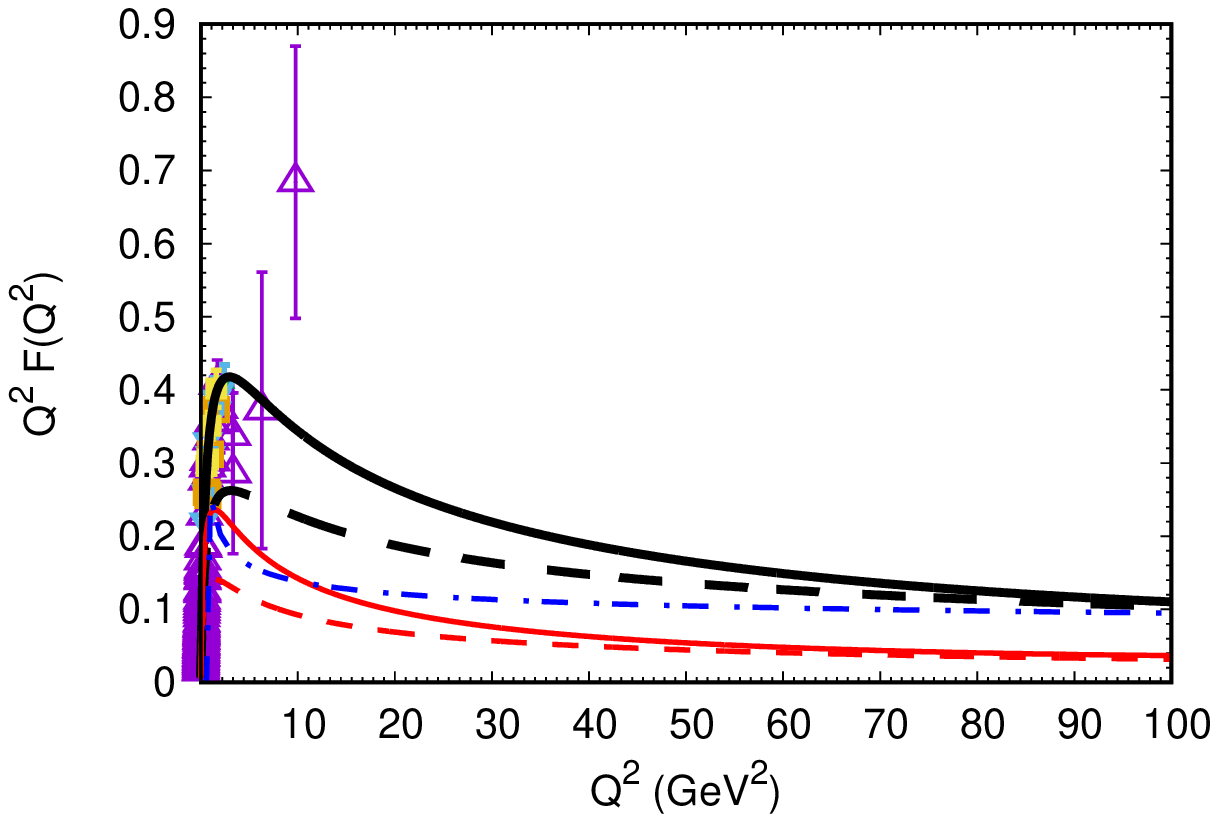}
  \caption{ Left panel: Pion em form factor and its LF valence contribution, both  divided by the monopole
   form factor $F_{mon}(Q^2)=1/(1+Q^2/m_\rho^2)$,   vs  $Q^2$.  Thick solid line: covariant form factor obtained by
   using the parameter set I in Table \ref{tableradius}, tuned
    to reproduce $f^{PDG}_\pi$.  
    Thick dashed line: the LF valence contribution to the FF calculated
   with the set I (notice that $F_{\rm val}(0)=P_{\rm val}$). Thin  lines: the same as the thick  ones, 
   but for the set II, 
   shown for the sake of comparison.  Experimental data correspond to: (i)   Refs.~\cite{baldini1999nucleon,baldini2000determination} (open upper triangle),  
  (ii)  Ref.~\cite{horn2006determination} (filled upper triangle),   (iii)  Ref. \cite{huber2008charged}
  (filled lower triangle), (iv)
   Ref.~\cite{tadevosyan2007determination} (open box) and (v)  Ref.~\cite{volmer2001measurement}
   (filled box).
  Right panel: the same as the left panel but with the  em FF  multiplied by $Q^2$,  for illustrating the asymptotic
  behavior.  Dot-dashed line: perturbative QCD prediction given by  Eq.~\eqref{pQCD}.
 } 
\label{FFvsexp}
\end{figure}

The pion em FF and its LF valence contribution divided by the monopole FF 
$F_{mon}(Q^2)=1/(1+Q^2/m_\rho^2)$  are  presented in the left panel of Fig.~\ref{FFvsexp} and    
compared 
with experimental data of Refs.~\cite{baldini1999nucleon,baldini2000determination,horn2006determination,huber2008charged,tadevosyan2007determination,volmer2001measurement}.
 The thick lines correspond to the calculations performed by using the parameter set I.
  Let us recall that this set has been tuned in order to reproduce  $f^{PDG}_\pi=130.50 (1)(3)(13)$ MeV
  \cite{Zyla:2020zbs} and, consequently,
  getting the charge radius in agreement  with $r^{PDG}_\pi$.  The thin lines
 have been obtained by using the parameter set II, that has been chosen in order to have an insight of the effects on the FF when varying the decay constant (see Table \ref{tableradius}).

 The pion em FF is  reproduced with  a remarkable accuracy by the  solution of the BSE obtained with the
 parameter set  I. Moreover,   the linear scale on  the ordinate axis allows one 
 to  observe that for  large enough momentum  $F_{\rm nval}(Q^2)$ 
  becomes subleading and dominated by $F_{\rm val}(Q^2)$, as  expected. 
Within  our dynamical model  (for both sets)  this valence dominance  occurs  above 
80 GeV$^2$, where  $F_{\rm val}(Q^2)/F(Q^2)$ starts to be $\geq~0.85$. 
 In the right panel of 
 Fig.~\ref{FFvsexp}, where a linear scale for the abscissa axis is adopted in order to emphasize the
behavior of the FF at  very
large values of $Q^2$, 
the FF and its LF valence contribution multiplied by $Q^2$ are shown.  Furthermore,
 for the sake of comparison, it is also presented  the QCD asymptotic FF,
 $F_{asy}(Q^2)$, given by Eq.~\eqref{pQCD} 
\cite{LepPLB1979}.
Our dynamical model becomes close to $F_{asy}(Q^2)$ only around
 $Q^2\sim 100\,\rm{GeV^2}$, and it is 
consistent with the previous analysis of the valence dominance in the pion form factor~\cite{LepPRD1980}.

 \begin{figure}[htp]
  \centering
  \includegraphics[scale=0.67, angle=0]{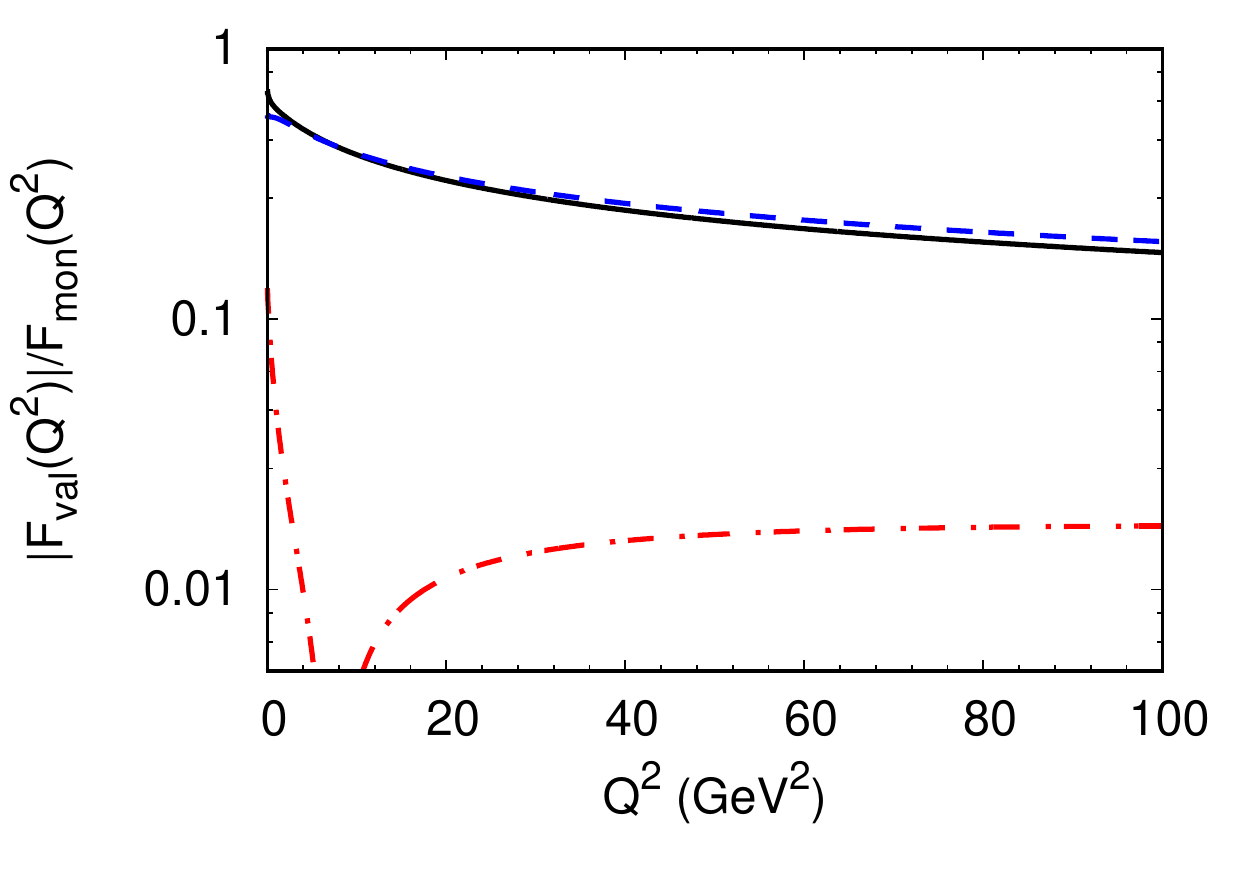} \hspace{-0.4 cm}
   \includegraphics[scale=0.67, angle=0]{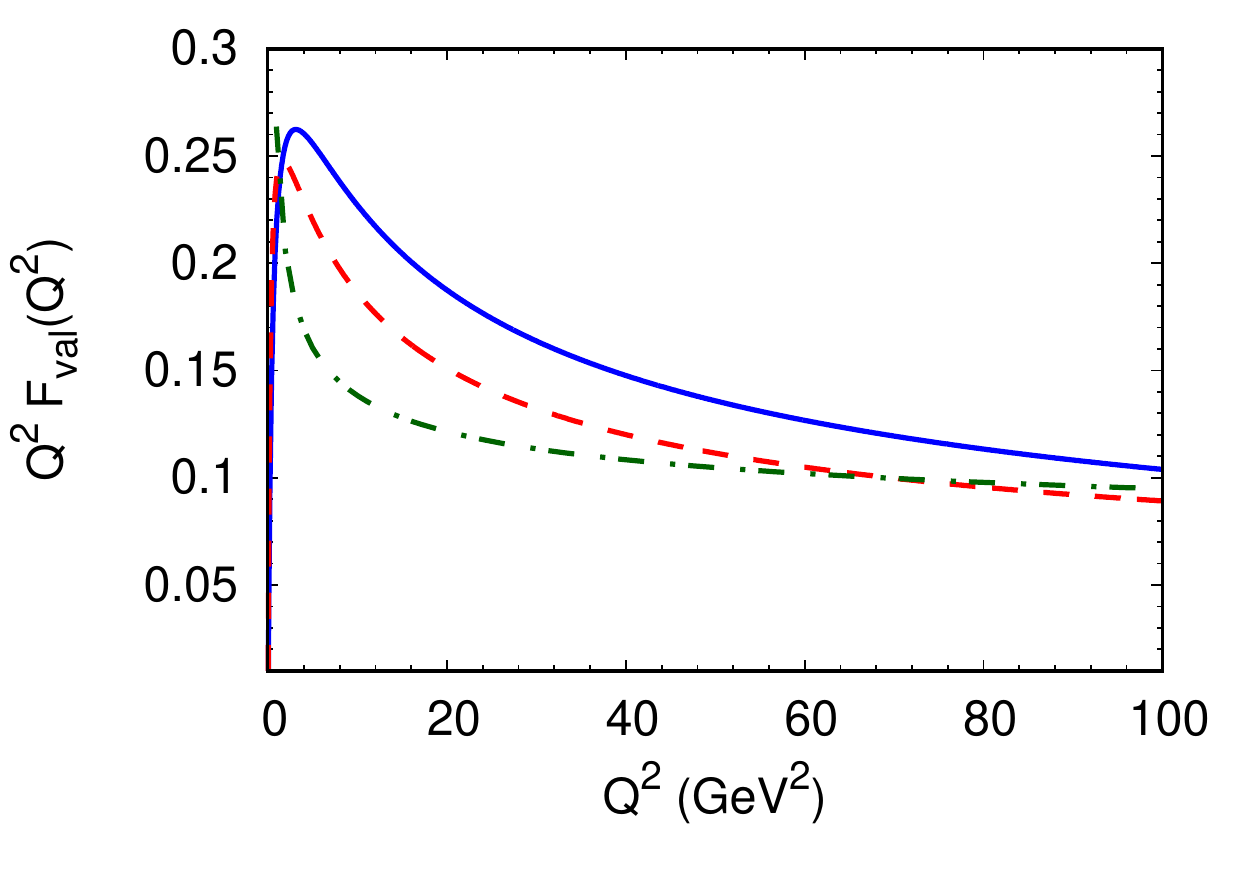}
  \caption{Left panel: anti-aligned (dashed line) and aligned (dot-dashed line) contributions to the valence 
  FF vs $Q^2$ compared with the total LF valence FF (solid line). All three FF's are divided by the monopole form factor.
 Right panel: asymptotic expression for the LF valence FF (dashed line), given in
 Eq.~\eqref{Eq:val_FF2}, compared to the valence FF (solid line), both multiplied by $Q^2$. 
 Results presented for set I. Dot-dashed line:  perturbative QCD prediction, Eq.~\eqref{pQCD}. 
 }
 \label{FFspin}
\end{figure}

 Within the BSE approach, it is  possible 
to evaluate even the contribution to the LF valence wave function from the two
different spin configurations present in the pion (see details in Ref.~\cite{dePaula:2020qna}). This
further decomposition of $F_{\rm val}$  is useful to
 illustrate the kinematical region where the purely relativistic aligned 
 configuration (i.e. $S=1$) 
  has interest. The   anti-aligned and aligned contributions to $F_{\rm val}$ 
  are
shown in the left panel of Fig.~\ref{FFspin}.
 Noteworthy, at $Q^2=0$  the aligned contribution is 
quantitatively appreciable, since   according  to  the probabilities of the $S=1$ and $S=0$ configurations~\cite{dePaula:2020qna} $F^{S=1}_{\rm val}(0)/F^{S=0}_{\rm val}(0)\sim 0.2$. 
For  increasing $Q^2$  the ratio $F^{S}_{\rm val}/F_{\rm mon}$  quickly decreases,  and $F^{S=1}_{\rm val}$ shows
   a dip below $Q^2=10~{\rm GeV^2}$. 
 The zero in the aligned FF is due to the  relativistic spin-orbit coupling that produces the term ${\bfm \kappa}\cdot {\bfm \kappa}'$ in the FF (see Eq.~\eqref{Eq:val_FF}), which flips the sign of this contribution around $Q^2\sim 8\,\rm{GeV^2}$. 
 In the right panel of Fig.~\ref{FFspin}, the approximate asymptotic expression 
 for the valence FF given in Eq.~\eqref{Eq:val_FF2} is
 compared  to the valence FF  and to the QCD  prediction given by Eq. \eqref{pQCD}. The differences
 between the results obtained from the dynamical model  are of the order of  20\%, at
 large $Q^2$,  while the full valence contribution is better approximated 
 by the perturbative QCD result.
 
\begin{figure}
    \centering
    \includegraphics[scale=0.67]{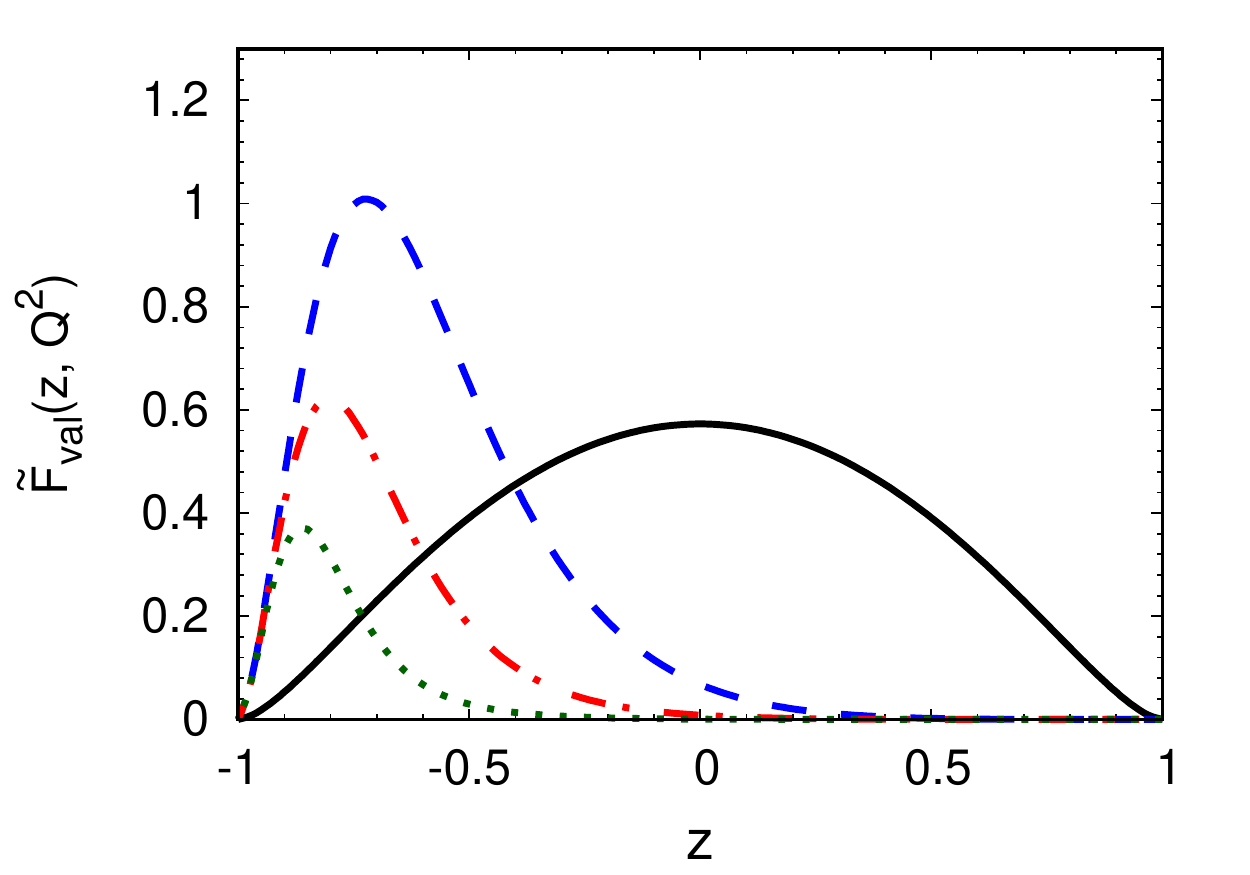} \hspace{-0.4 cm}
    \includegraphics[scale=0.67]{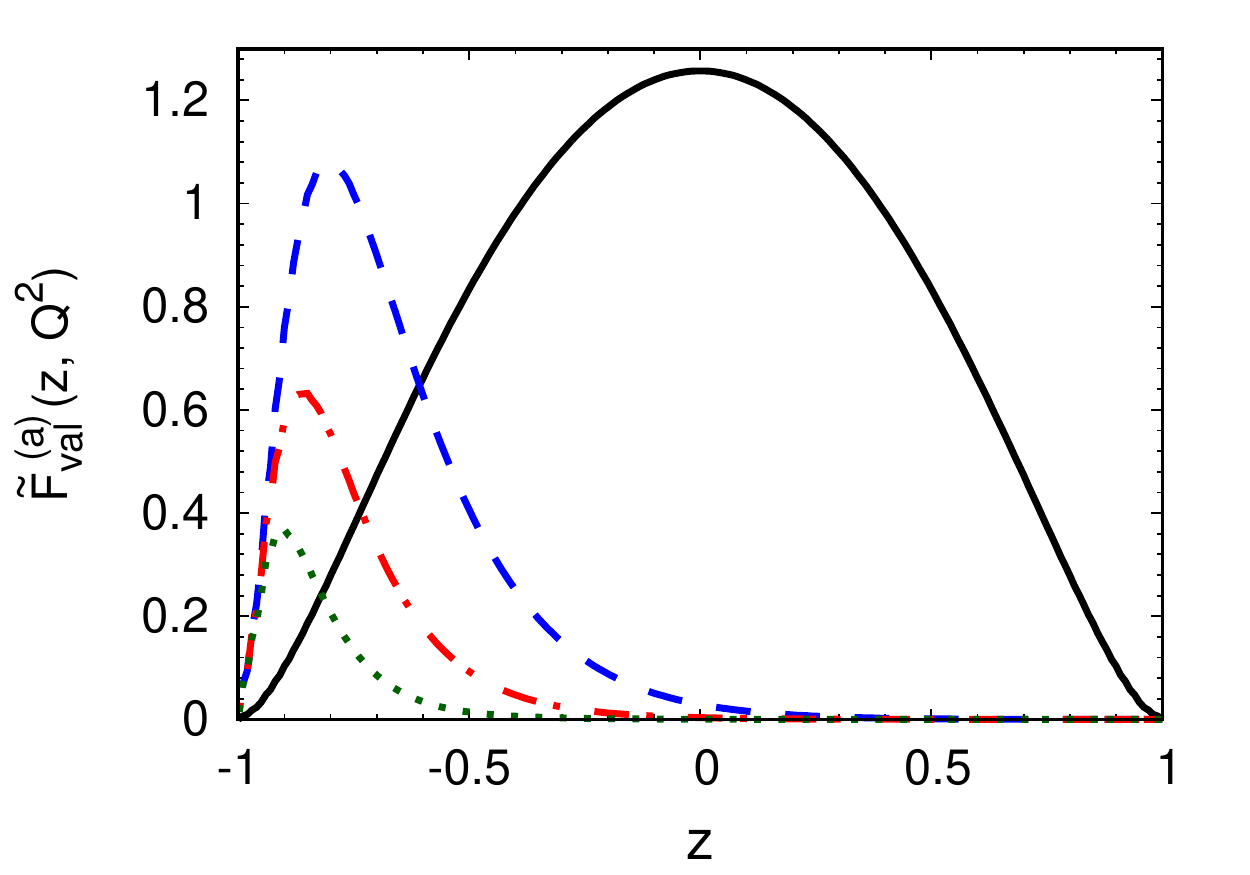}
    \caption{Left panel: $\tilde F_{\rm val}(z,Q^2)$  vs $z$ and  fixed values of $Q^2$. Solid line: $Q^2=0$.  Dashed line:      $Q^2=5~{\rm GeV^2}$.      Dashed-dotted line: $Q^2=10~{\rm GeV^2}$.  Dotted line: $Q^2=20~{\rm GeV^2}$. For the sake of visibility the results for $Q^2 > 0$ have 
    been multiplied by a factor of 10.  
      Right panel:
     the same as the left one, but for $\tilde F^{(a)}_{\rm val}(z,Q^2)$, 
    (cf.
    Eq.~\eqref{Eq:val_FF2}). The results are presented for set I in both panels.}
    \label{FFvsxi}
\end{figure}

  A 3D imaging
 of the pion charge distribution could be reached by a further analysis to be performed within 
 our dynamical approach~\cite{dePaula:2021}. As a matter of fact, the covariant FF can be written  
   in the $q^+=0$ frame as follows 
 (see, e.g., Refs.~\cite{Soper:1976jc,Brodsky:2007hb,Miller:2007uy})) 
 \begin{equation}
 F(Q^2) = \int_{-1}^1 {dz\over 4\pi}\int {d{\bf b_\perp}\over (2\pi)^2}~ e^{-i{\bf q}_\perp\cdot {\bf b}_\perp}~
 \rho_{ch}\Bigl(z,|{\bf  b}_\perp|\Bigr)=\int_{-1}^1 dz~ \tilde F(z,Q^2), 
 \end{equation}
 where ${\bf b}_\perp$ is  
 a vector in the impact parameter space, and $\rho_{ch}\Bigl(z,|{\bf  b}_\perp|\Bigr)$  is the charge density and $\tilde F(z,Q^2)$ a new
 quantity that we call  {\em sliced form factor}.   Noteworthy,  such a quantity, for given
 $z$ and ${\bf b}_\perp$, is
 the  no spin-flip generalized parton distribution with a vanishing skewness (see, e.g.,
 Ref. \cite{Frederico:2009fk}).

In the present work, we focus only on the valence component,  leaving   the sliced full  FF and 
$\rho_{ch}\Bigl(z,|{\bf  b}_\perp|\Bigr)$ to a detailed analysis in a future work.
 In Fig.~\ref{FFvsxi}, the $z$-dependence of the sliced valence FF 
 is analyzed through two quantities:  (i) 
$\tilde
F_{\rm val}(z,Q^2)$; and (ii) $\tilde F^{(a)}_{\rm val}(z,Q^2)$,
 obtained from the  asymptotic form given by Eq.~\eqref{Eq:val_FF2}. The analysis is done for $Q^2=0,~5,~10 ~ \mbox{and}~20 ~{\rm GeV^2}$.
 In the left panel, the shape of   $ \tilde
F_{\rm val}(z,Q^2)$ supports
  the dominance of the end-point regions  
  for large $Q^2$, as explained below Eq.~\eqref{Eq:val_FF2}. 
  As a
  matter of fact, at $Q^2=0$ one has a symmetric behavior, while for increasing values of $Q^2$ the sliced FF's 
  sizably cumulates
  in the region close to $z=-1$ for a quark,  as the $q\bar q$ pair is collinear to preserve the pion in the final state.
   Notice that the support of the bump  shrinks as $\sim Q^{-2}$,  which can be inferred by examining the approximate asymptotic  formula given by Eq.~\eqref{Eq:val_FF2}.  

%%%%%%%%%%%%%%%%%%%%%%%

{\it Summary.} 
We developed a fully  four-dimensional treatment
 of the pion electromagnetic form factor based on the actual solution of the  Bethe-Salpeter
 equation in Minkowski space. The ladder kernel is considered,  as suggested by
 the suppression of the non-planar contributions for $N_c=3$ within  the BS approach  in a scalar QCD model~\cite{Nogueira:2017pmj}. The three inputs parameters, namely (i) the constituent mass, (ii) the gluon 
 mass and (iii) the scale of the
quark-gluon vertex,  are inspired by the infrared properties of QCD, as obtained by LQCD.  Once a fine
tuning is carried out  to reproduce  
the decay constant $f^{PDG}_\pi$, 
the obtained charge radius, $r_\pi= 0.663~\text{fm}$,  nicely agrees  with the most recent experimental value, $r^{PDG}_\pi= 0.659 \pm 0.004~\text{fm}$ \cite{Zyla:2020zbs}.  In correspondence, the 
  predicted  valence and 
non-valence radii amount to  0.71~fm and 0.54~fm, respectively. This means 
that the charge distribution generated by the  $q\bar q$ pair in the higher Fock-components of the
pion LF wave function is considerably more compact ($\sim$25\%) than the valence 
configuration, although still significant. One should
also keep in mind that 
 the valence probability is about 70\% and
the remaining probability is associated with the occupancy of states with a $q\bar q$ pair and any number 
of gluons. The general behavior of the calculated em FF is illustrated in Fig.~\ref{FFvsexp} and it is found to be  in remarkable
agreement with  the existing
data for  $Q^2\leq 10~{\rm GeV^2}$. It is also interesting to notice that our covariant calculation matches the well-known perturbative QCD expression \eqref{pQCD} for $Q^2>90~{\rm GeV^2}$, namely in the kinematical
 region where the valence  contribution becomes dominant. Two further analyses have been presented, that
 allow a deeper understanding of the dynamics inside the pion. First, the decomposition of the valence
 component of the FF in terms of the  spin configurations $S=0$,  with about 75\% of the
 valence probability, and $S=1$.  The aligned spin configuration is a purely
 relativistic effect and  has  an impact at $Q^2=0$, counterintuitively (though understandable  by 
 the normalization constraint), but it becomes negligible at large $Q^2$. The second analysis  is given by the study of the influence of the end-point behavior, i.e. the region
 close to $z=\pm 1$, on the  valence FF at different values of the momentum transfer. Such an initial
 investigation could
 open a more stringent analysis of the charge distribution in the impact parameter space to explore the pion structure. 

The unique features of  Minkowski space methods applied to the continuum 
QCD approach are of  great importance and need to be further explored. Future plans are to include (i) dressing functions 
 for quark and gluon propagators (see Ref.~\cite{Mezrag:2020iuo}, for an extension of
the NIR
framework to the fermion-photon coupled  gap equations), and (ii) a more 
realistic quark-gluon vertex. The IR properties of the form factor that we have quantitatively explored 
 within the Minkowski space approach might inspire further experimental works to disentangle 
 valence and beyond-valence contributions.

%\bigskip

\textit{Acknowledgments.} J.H.A.N.~  gratefully thanks C\'edric Mezrag for helpful discussions. This study was financed in part by Conselho Nacional de Desenvolvimento Cient\'{i}fico e Tecnol\'{o}gico (CNPq) under the grant 438562/2018-6 and 313236/2018-6 (WP), and 308486/2015-3 (TF), and 
by Coordena\c{c}\~ao de Aperfei\c{c}oamento de Pessoal de N\'{i}vel Superior (CAPES) under the grant 88881.309870/2018-01 (WP).
J.H.A.N.~acknowledges the support of the grants \#2014/19094-8 from Funda\c{c}\~ao de Amparo \`{a} Pesquisa do Estado de S\~ao Paulo (FAPESP). E.Y.~thanks for the financial
support of the grants \#2016/25143-7 and \#2018/21758-2 from FAPESP.  We  thank the FAPESP Thematic Projects
grants   \#13/26258-4 and \#17/05660-0.  

\appendix
\section{Coefficients}
\label{App:4Di}
 The coefficients $ \tilde c_{ij}$ in Eq. \eqref{Eq:FF_final} are 
 given by   
\be
    \tilde c_{11} = \tilde c_{22} = -6\Bigl(\alpha + {1\over 2}\Bigr)~,
     \quad \tilde c_{12} =\tilde  c_{21} = 6
    \frac{m}{M}~, 
    \quad \tilde c_{13} =\tilde  c_{24} = - 3 {m\over M}~ {Q^2\over M^2}~\beta~, \nonu
   \tilde  c_{14} = \tilde c_{23} = 3~{M^2_{\rm{cov}}\over M^2} - 3 ~ {Q^2\over 2M^2}~(2 \alpha+1-4\beta)~\beta~, 
   \quad \tilde c_{31} = \tilde c_{42} = -3 {m\over M}~{  Q^2\over 2M^2}~(2 \alpha+1-2\beta)~, \nonu
   \tilde  c_{32} = \tilde c_{41} = 3~{M^2_{\rm{cov}}\over M^2} + 3 {Q^2\over 4M^2}~\Bigl[(2\alpha+1)^2-12\beta \alpha
   +8\beta^2 -6 \beta\Bigr]~, \nonu
   \tilde  c_{33} = \tilde c_{44} = - {M^2_{\rm{cov}}\over M^2}~ (2\alpha +1)~\Bigl({3\over 2} +  {Q^2\over M^2}\Bigr)
   +3 ~{Q^2\over 2M^2} ~\Bigl(1+{Q^2\over 4M^2}\Bigr)~(2\alpha +1)~(2 \alpha+1-2\beta) \beta
   ~, \nonu
   \tilde  c_{34} = \tilde c_{43} = - {m\over M} ~{M^2_{\rm{cov}}\over M^2}~\Bigl (3 + {Q^2\over 2 M^2}\Bigr) 
   + 3{m\over M}~  {Q^2\over M^2}~ \Bigl(1+{Q^2\over 4M^2}\Bigr)~~(2 \alpha+1-2\beta)~\beta~~.
\ee
where 
$\alpha =[z \, y + z' \, (1-y)] /2$ and
$ \beta = (1-y) ~ (1+z') / 2$.

%\bibstyle{elsarticle-num-names}
%\bibstyle{numcompress}
%\bibliography{pionff} 

\end{document}